\documentclass[aps,prl,twocolumn,showpacs,10pt,superscriptaddress,preprintnumbers,nofootinbib,longbibliography]{revtex4-1}
\usepackage{epsfig,amssymb,amsmath,psfrag,epstopdf,color,siunitx}
\pdfoutput=1
\usepackage{graphicx}
\usepackage[caption=false]{subfig}
 \usepackage[export]{adjustbox}
 \usepackage{tikz}
\newlength{\figthreewidth}
\usepackage{paralist}
\usepackage{hyperref}
\usepackage[normalem]{ulem}
\allowdisplaybreaks

\newcommand{\nn}{\nonumber}

\DeclareRobustCommand{\Fig}[1]{Fig.~\ref{#1}}

\DeclareRobustCommand{\Eq}[1]{Eq.~(\ref{#1})}

\begin{document}
\title{Observing Macroscopic Consequences of Electroweak Anomalies\\ with Archival DELPHI Data}

\author{Jingyu Zhang}
\email{jingyu.zhang@cern.ch}
\affiliation{Department of Physics and Astronomy, Vanderbilt University, Nashville, TN, USA}

\author{Kyle Lee}
\email{kyle@anl.gov}
\affiliation{High Energy Physics Division, Argonne National Laboratory, Lemont, IL, USA}

\author{Ian Moult}
\email{ian.moult@yale.edu}
\affiliation{Department of Physics, Yale University, New Haven, CT 06511, USA}

\author{Yi Chen}
\email{luna.chen@vanderbilt.edu}
\affiliation{Department of Physics and Astronomy, Vanderbilt University, Nashville, TN, USA}

\author{Yen-Jie Lee}
\email{yenjie@mit.edu}
\affiliation{Laboratory for Nuclear Science, Massachusetts Institute of Technology, Cambridge, Massachusetts, USA}

\begin{abstract}

In this \emph{Letter}, we emphasize that asymmetries in charge flux produced in the decays of on-shell Z-bosons provide a macroscopic manifestation of electroweak anomalies in the Standard Model (SM). 
We propose that these can be cleanly observed using charge correlators, providing a new formulation of forward-backward asymmetry measurements that is particularly well suited for precision studies of hadronic decays.
Using archival DELPHI data, we perform a first measurement of the one-point charge correlator of electromagnetic charge flux on hadrons, and cleanly observe the macroscopic imprint of the underlying anomaly. 
Our analysis illustrates the potential of charge correlators as precision electroweak observables, and motivates a renewed effort to resolve longstanding tensions in hadronic asymmetry measurements.

\end{abstract}

\maketitle

\section{Introduction}

Anomalies play a central role in our understanding of quantum field theory: trace anomalies govern our understanding of the scale dependence of quantum field theories (QFTs) in even dimensions via the celebrated $a$ and $c$ theorems \cite{Zamolodchikov:1986gt,Cardy:1988cwa,Komargodski:2011vj}; 't Hooft anomalies constrain renormalization group flows and infrared phases of QFTs \cite{tHooft:1979rat}; (the absence of) gauge anomalies constrain the matter content, representations, and space-time dimensionality of consistent gauge  \cite{Weinberg:1967tq,Georgi:1974sy,Witten:1982fp} and string theories \cite{Green:1984sg,Polyakov:1981rd}.  Beyond their theoretical interest, anomalies are responsible for numerous remarkable macroscopic phenomena realizable in the laboratory, including negative magneto-resistance \cite{xiong2015evidence,huang2015observation}, anomalous thermoelectric transport \cite{Gooth:2017mbd}, and the chiral magnetic effect \cite{Li:2014bha}.

The SM of particle physics possesses a rich set of (approximate) global symmetries, many of which exhibit mixed 't Hooft anomalies with the electroweak $SU(2)_L \times U(1)_Y$ symmetries, 
\begin{align}
\hspace{-0.2cm}\partial_\mu J_Q^\mu=\frac{1}{16 \pi^2}\left[d^{ab}_{W\!W\!Q}\, W_{\mu \nu}^a \widetilde{W}^{b \mu \nu}+d_{Y\!Y\!Q}\, Y_{\mu \nu} \widetilde{Y}^{\mu \nu}\right]\,.
\label{eq:Banom}
\end{align}
Here $Q$ denotes a generic charge, for example, baryon number, lepton number, or a flavor symmetry. 
Due to great experimental effort, the couplings and quantum numbers of the SM are (mostly) well measured, making the calculation of the anomaly coefficients in \Eq{eq:Banom} a textbook exercise \cite{Peskin:1995ev}. 
However, macroscopic phenomena manifesting the effects of \Eq{eq:Banom} in terrestrial experiments are harder to come by:
proposed possibilities include anomaly-mediated neutrino-photon scattering \cite{Harvey:2007ca,Harvey:2007rd}, and the production of non-trivial topological field configurations at colliders \cite{Khlebnikov:1990ue,McLerran:1989ab,Espinosa:1989qn,Ringwald:1989ee,Gibbs:1994cw,Ringwald:2002sw,Bezrukov:2003er,Bezrukov:2003qm,CMS:2018ozv,Ellis:2016ast,Ellis:2016dgb,Tye:2015tva,Khoze:2019jta,Khoze:2020tpp,Matchev:2025ivr}.
Apart from the general interest in testing structural aspects of quantum field theory, phenomena mediated by anomalies are often robust, even in messy real-world experiments, making them ideal candidates for precision SM measurements. 

\begin{figure}[t]
\centering
\includegraphics[width=0.34\textwidth]{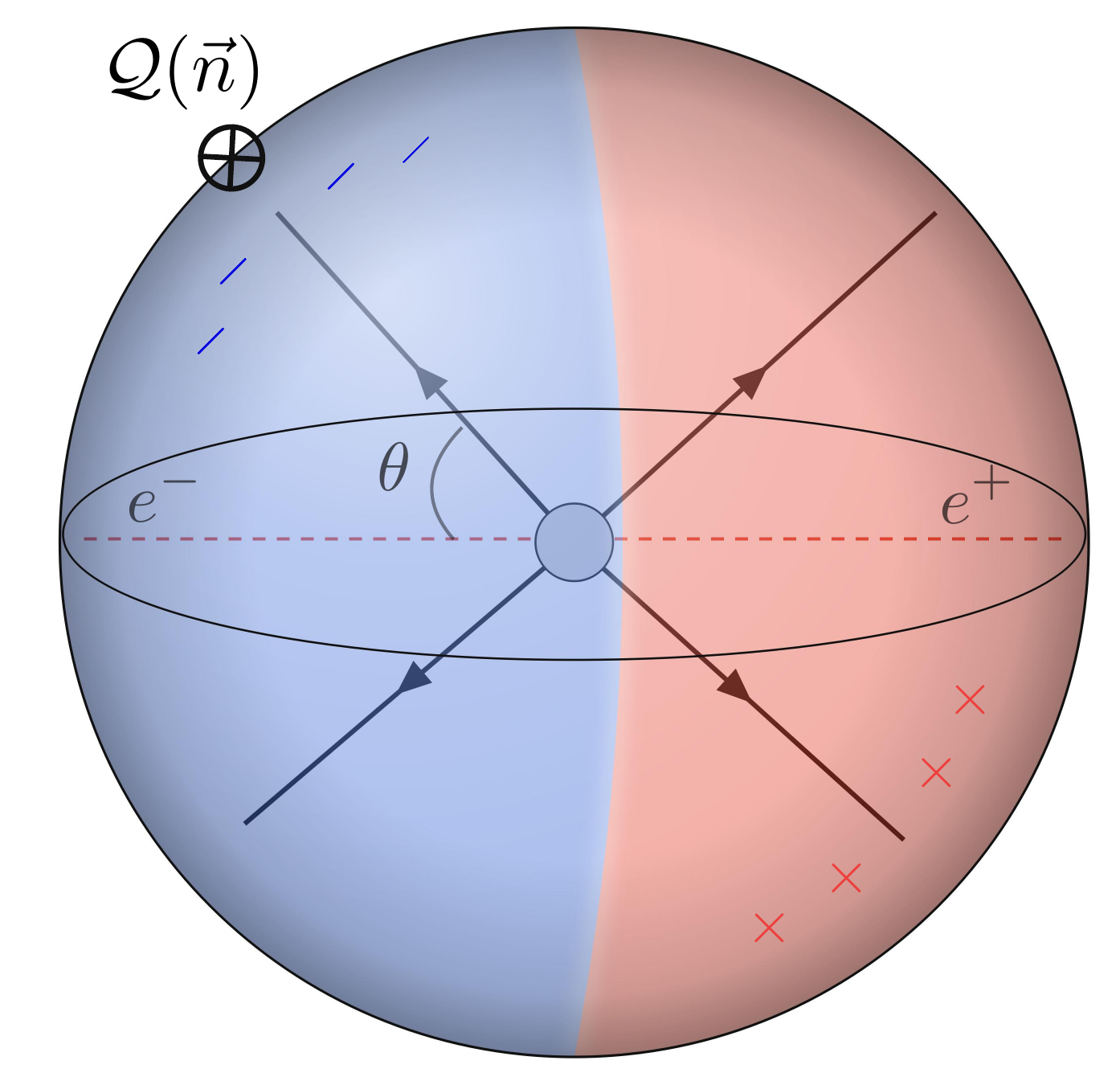}
\caption{Decays of Z-bosons produced in electron-positron collisions exhibit a flux asymmetry related to the presence of a mixed electroweak anomaly of the current being measured.}
\label{fig:anomaly_detector_a}
\end{figure}

In this \emph{Letter}, we emphasize a clean macroscopic consequence of \Eq{eq:Banom}: asymmetries in charge fluxes of decaying on-shell Z bosons, as produced in high statistics at LEP and SLC.   
We show that if these fluxes are measured with charge correlators, see \Fig{fig:anomaly_detector_a}, the asymmetry can be directly related to the anomaly of the underlying theory.
Our framing provides a new set of observables, and a fresh perspective, on the well-studied physics of forward-backward asymmetries at the Z-pole \cite{ALEPH:2005ab,ALEPH:2010aa}. 
Due to its relation to the protected anomaly, our perspective is particularly powerful for many-body hadronic states, where tensions in forward-backward asymmetry measurements have persisted for many years \cite{Baak:2014ora,Haller:2018nnx,Fischer:2026bka}.
Using archival data from the DELPHI experiment~\cite{DELPHI:1990cdc, DELPHI:1995dsm}, we perform the first measurement of these correlations, illustrating in real data that they provide a clean macroscopic illustration of the underlying anomaly, and providing a first indication of their potential for precision physics.

\section{Collider Fluxes and Anomalies}

Although it is challenging to turn on the $SU(2)_L \times U(1)_Y$ background fields in \Eq{eq:Banom}, collider experiments produce electroweak $W$ and $Z$ bosons at short distances, allowing access to the consequences of the anomaly in the linear response regime.
In the simplified context of hypothetical collider experiments in conformal field theories (CFTs), Ref.~\cite{Hofman:2008ar} showed how anomalies of the currents coupling to particles produced at intermediate stages of a high-energy collision are imprinted in asymptotic fluxes.
We review this in the context of CFTs, before showing how it can be generalized to the SM, and realized experimentally using high-statistics samples of on-shell Z-bosons produced at LEP and SLC.

The approach of \cite{Hofman:2008ar} uses the canonical signature of an anomalous current, namely the presence of contact or ``Schwinger" terms  \cite{Schwinger:1959xd} in the divergence of an anomalous three-point function of currents \cite{Adler:1969gk,Bell:1969ts,Adler:1969ccs,Bardeen:1969md,Wilson:1969zs}
\begin{align}\label{eq:contact}
\hspace{-0.2cm}\partial_\mu^x \langle J^\mu_Q(x) J^\nu (y) J^\rho(z) \rangle \\
&\hspace{-1.4cm}\propto d_{J\!J Q}\epsilon^{\nu\rho\alpha\beta}\partial^y_\alpha \partial^z_\beta \delta^{(4)}(x-y) \delta^{(4)}(x-z)\,,\nonumber
\end{align}
as appears in the familiar $\pi^0\to \gamma \gamma$ decay \cite{Adler:1969gk,Bell:1969ts}. In a CFT, this contact term, and hence the anomaly, uniquely fixes the structure of the full three-point function, $\langle J^\mu_Q(x) J^\nu (y) J^\rho(z) \rangle$ \cite{Schreier:1971um,Osborn:1993cr}.
While the three-point function itself cannot be directly measured in colliders, the flux of a U$(1)$ charge
\cite{Hofman:2008ar}
\begin{align}
\label{eq:chargedet}
\mathcal{Q}(\vec n) = \lim_{r\to\infty} r^2 \int_{0}^\infty dt\, n_i\, J^i_Q(t, r\vec n)\,,
\end{align}
as a function of the angle on the detector can be measured in momentum eigenstates 
\begin{align}
J=\int d^4 x\, e^{i q\cdot x} J(x)\,,
\end{align}
where $q=(E,0,0,0)$. Since we always work with momentum eigenstates, we drop the momentum label for simplicity. To condense notation, operators without explicit position dependence are Fourier transformed in this manner.
This expectation value,
$\langle  \mathcal{Q}(\vec{n})\rangle_{J}\equiv \langle 0| J^\dagger \mathcal{Q}(\vec{n}) J |0\rangle /\langle 0| J^\dagger  J |0\rangle $,
can be directly computed from the current three-point function, and takes the form \cite{Hofman:2008ar}
\begin{align}
\langle  \mathcal{Q}(\vec{n})\rangle_{J}&
= \frac{3}{\pi}\frac{d_{J\! J Q}}{c_{J}}\frac{i\epsilon^{ijk} \epsilon^*_i \epsilon_j n_k}{\epsilon^*_l \epsilon_l}\,,  
\label{eq:oneptCharge}
\end{align}
where $\epsilon$ is the polarization vector of the current.
Expressed in terms of lab variables, this manifests as a characteristic $\cos(\theta)$ asymmetry, where $\theta$ is the angle between the spin vector and the direction of the charge detector.

\begin{figure}[t]
\centering
\includegraphics[width=0.49\textwidth]{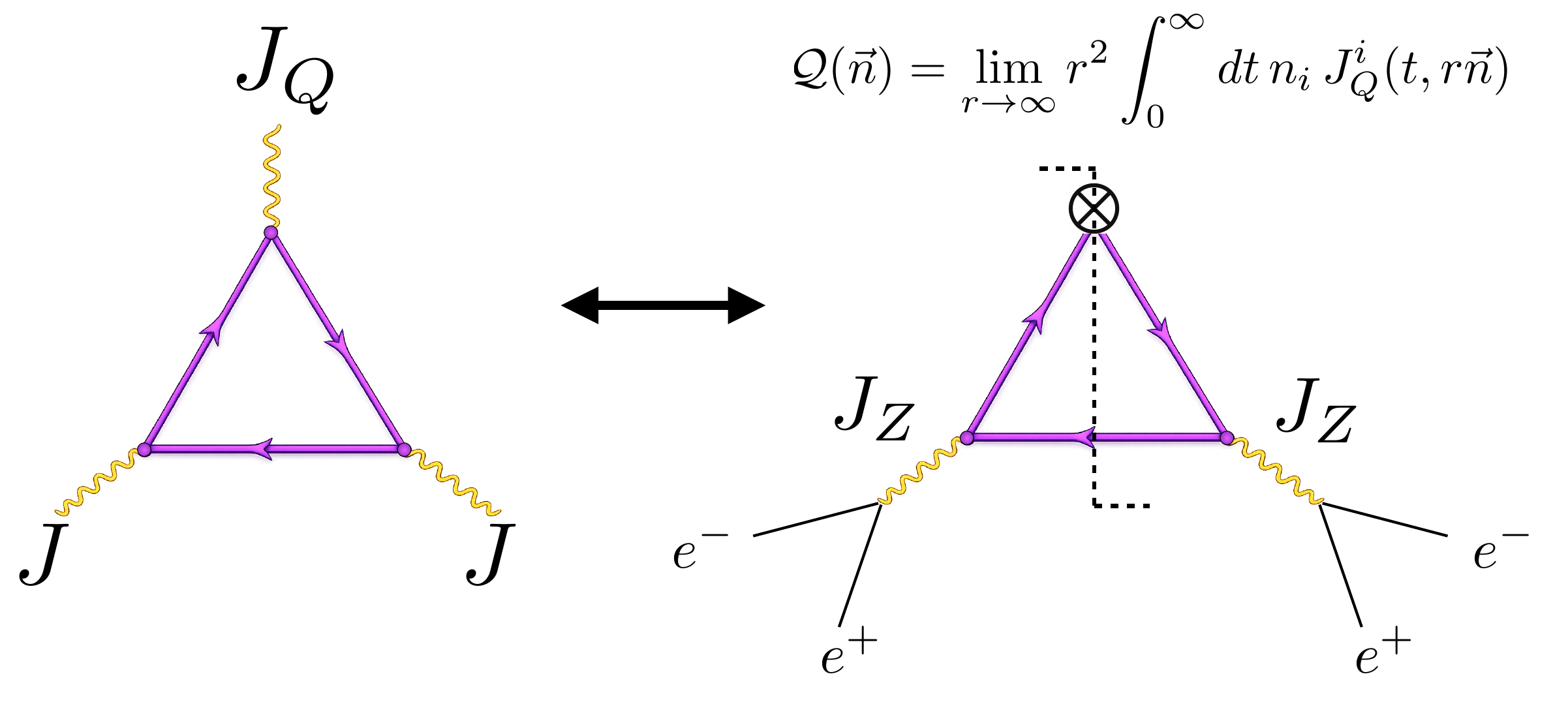}
\caption{An anomalous $\langle J_QJJ \rangle$  three-point function can be realized experimentally by tuning $s=m_Z^2$ in an $e^+e^-$ collider, and measuring the flux of an anomalous current, $\langle  \mathcal{Q}(\vec{n})\rangle$.}
\label{fig:anomaly_detector}
\end{figure}

This direct relation between the anomaly and the charge correlator is illustrated in \Fig{fig:anomaly_detector}, which also emphasizes the well-known fact that the contact term for a local anomaly is fixed by long distance effects \cite{Dolgov:1971ri,Coleman:1982yg,Frishman:1980dq}. For a recent study of anomalies from an on-shell perspective, see \cite{Kakkad:2026ynv}.

We therefore see in the simplified setting of ``conformal colliders" how anomalies manifest in macroscopic fluxes.
Conformal colliders directly probe the non-conservation of an anomalous current without necessitating the generation of topologically non-trivial field configurations.  Unlike the total charge $\mathcal{Q} =\int d\Omega\, \mathcal{Q}(\vec n)$, $\mathcal{Q}(\vec n)$ is a non-topological symmetry operator, related to large gauge transformations of the background U$(1)$ field \cite{Strominger:2017zoo}. While $\langle  \mathcal{Q} \rangle_{J}=0$, $ \langle\mathcal{Q}(\hat n)\rangle_{J}$ is non-zero even in topologically trivial field configurations, allowing for a macroscopic consequence of the anomaly without net charge production.

This approach is not specific to charge. Indeed, it applies equally well to the more familiar case of energy flux \cite{Hofman:2008ar}, but unfortunately not for the states easily accessed in real-world colliders. In analogy with Eq. \ref{eq:Banom}, in the presence of a background gravitational field, a four-dimensional conformal field theory exhibits a trace anomaly \cite{Duff:1993wm,Deser:1976yx,Capper:1974ic}
\begin{align}\label{eq:T_anom}
T^\mu_\mu=\frac{c}{16 \pi^2} W^2 -\frac{a}{16 \pi^2}E\,,
\end{align}
where $W$ is the Weyl tensor, $E$ is the Euler density, and $a$ and $c$ are anomaly coefficients.  To slightly simplify the presentation, we restrict ourselves to an $\mathcal{N}=1$ superconformal field theory. In this case, much in analogy with the $\langle JJJ\rangle $ three-point function, the $\langle TTT \rangle$ three-point function is completely fixed by the anomaly coefficients \cite{Osborn:1998qu,Erdmenger:1996yc,Osborn:1993cr}. The flux of energy  \cite{Sterman:1975xv,Korchemsky:1997sy}
\begin{align}
\mathcal{E}(\vec n) = \lim_{r\to\infty} r^2 \int_0^\infty dt\, n^i\, T_{0i}(t, r\vec n)\,,
\end{align}
in a stress tensor state, $\langle 0| T \mathcal{E}(\vec n)T |0\rangle/\langle 0| T T |0\rangle \equiv \langle \mathcal{E}(\hat n) \rangle_T$ is then also completely fixed in terms of the anomaly \cite{Hofman:2008ar},
\begin{align}
\langle \mathcal{E}(\hat n) \rangle_T=\frac{\mathcal{E}}{4\pi}\left[1+\frac{6(c-a)}{c} \left(\frac{\epsilon^*_{ij} \epsilon_{il} n_j n_l}{\epsilon^*_{ij} \epsilon_{ij} } -\frac{1}{3} \right) \right]\,.
\end{align}

In exact analogy with the case of charge flux, while the total energy $\mathcal{E}=\int d \Omega\,\mathcal{E}(\vec n)$ is topological, $\mathcal{E}(\vec n)$ is a non-topological symmetry generator of supertranslations of the Bondi-Meisner-Sachs group \cite{Strominger:2017zoo}. Mixed gravitational anomalies can be studied in a similar manner \cite{Cordova:2017zej, Meltzer:2017rtf}.

We note that there is a long history of studies of the trace anomaly in QCD, see e.g. \cite{Novikov:1980fa,Voloshin:1980zf}. These focus on accessing the contribution to the trace anomaly from background gauge fields, as opposed to gravitational fields in \Eq{eq:T_anom}, via the $\langle J T J \rangle$ three-point function.  This can be done in deeply virtual Compton scattering (where they are referred to as gravitational form factors \cite{Burkert:2023wzr}) or in polarized Raman scattering \cite{Nguyen:2021vrb}, and can also be analyzed using the techniques of conformal field theory \cite{Coriano:2026gyl,Coriano:2024qbr,Coriano:2011ti}. The novelty of this approach is the relation between local correlators and detector operators, which allows their study using asymptotic fluxes in colliders. 

The elegance of this relation motivates exploring whether it can be realized in the SM, and more importantly, if it can be used to improve precision SM measurements.
A key simplification of the conformal collider setup is the use of a local operator to produce the collider state, whereas real-world colliders use beams of particles. 
Starting from the earliest studies of $e^+e^-$ colliders \cite{Cabibbo:1961sz,Putzolu:1961jya}, it was known that asymmetries proportional to $\cos(\theta)$, as in \Eq{eq:oneptCharge}, are produced by interference terms between radiative corrections even in pure QED.
Indeed, the need to understand this background source of asymmetry motivated the original calculations of one-loop radiative corrections to electron-positron scattering in QED \cite{FURLAN1964262,Khriplovich:1973by,Dicus:1973jh,Brown:1973ji,Berends:1973fd} and in the electroweak theory \cite{Passarino:1978jh,Bohm:1983rn} (for the complete two-loop result in QED, see  \cite{Gerasimov:2026zmh}).
For this reason, $\cos(\theta)$ terms in angular distributions are historically associated with interference phenomena, and are indeed not generically associated with anomalies.

However, the conformal collider setup \emph{can} be achieved experimentally when electron-positron colliders are precisely tuned to the Z-pole. In this case, interference effects are suppressed by the Z-boson width, $\Gamma_Z/m_Z$, 
providing nearly pure access to correlators of the chiral spin-1 current, 
\begin{align}
J_Z^\mu = c_W^2  J_{T_3}^\mu -s_W^2  J_Y^\mu\,.
\end{align}
Here $J_{T_3}^\mu$ is the third-component of the $SU(2)_L$ current, $J_Y^\mu$ is the hypercharge current, and $c_W = \cos\theta_W$, $s_W = \sin\theta_W$, with the Weinberg angle $\theta_W$.  
In this special case of collisions on the Z-pole, the measurement of any detector formed from a U$(1)$ charge that has mixed anomalies with $J_Z$ (i.e. with $J_{T_3}$ or $J_Y$), provides direct experimental access to the physics of the anomaly, which manifests as the characteristic $\cos(\theta)$ angular distribution of \Eq{eq:oneptCharge} in macroscopic charge fluxes.

In this \emph{Letter}, we extend the relation between anomalies and charge correlators beyond the case of conformal field theories to the real-world SM, and realize it using precision measurements of charge correlators at DELPHI.

\section{Anomalies in Leptonic Fluxes and Forward-Backward Asymmetries}

Although our formulation of charge fluxes in terms of detector operators is most powerful for high-multiplicity hadronic events, we begin by illustrating it in the simplest case of leptonic final states. This allows us to explain its relation to classic studies of forward-backward asymmetries in a simplified setting, before generalizing to more complex final states. We emphasize that angular distributions of leptons are among the earliest observables measured in electron-positron colliders \cite{Bernardini:1974gx,Borgia:1972ny}, and leptonic asymmetries in particular are responsible for much of our understanding of the electroweak sector of the SM \cite{JADE:1981jmg,Adeva:1982pm,TASSO:1982oyp,CELLO:1982tjl,Fernandez:1983ua,Levi:1983fd,TASSO:1983fzg,PLUTO:1983fem,DELPHI:1999yep,ALEPH:1997gvm,L3:1995nbn,L3:1997vix,OPAL:1997hce}. In leptonic final states, it is not necessary to use the framework of detector correlators. Nevertheless, we use them as a pedagogical example, before continuing to the case of hadronic final states, where we believe that our approach offers an advantage over previous formulations.
 
 The simplest example of a $U(1)$ current we can consider  is $J_{\gamma_l}=\bar l \gamma^\mu l$, where $l=e,\mu, \tau$. In a collider experiment, the detector operator in \Eq{eq:chargedet} can then be interpreted as measuring the electromagnetic charge restricted to leptons in a single generation, hence the notation $J_{\gamma_l}$. This current has a mixed electroweak anomaly in the SM. This is easy to understand, since $J_{\gamma_l}$ is simply the restriction of electromagnetic charge to the lepton sector. Since anomaly cancellation for the electromagnetic gauge current in the SM famously occurs between leptons and quarks \cite{Gross:1972pv,Bouchiat:1972iq}, the restriction to the leptonic sector is anomalous. The anomaly coefficient with two $Z$ currents and $J_{\gamma_l}$ can be written as a linear combination of $T_3$, $Y$ and $L$ as
 \begin{align}
\label{eq:dZZgammaL}
\hspace{-0.3cm}d_{ZZ\gamma_l} &=-\frac{1-4 s_W^2}{2} d_{T_3 T_3 L}+s_W^4\left(d_{T_3 T_3 L}+d_{Y Y L}\right)\,.
\end{align} 
Although the SM is not conformal, for a leptonic decay of a Z-boson it is a good approximation to treat the lepton as massless, and ignore electroweak radiative corrections (We will soon generalize to incorporate the breaking of conformal symmetry perturbatively). In this approximation we have conformal symmetry, allowing us to use the relation between the anomaly coefficient
and the charge flux distribution in a conformal field theory (\Eq{eq:oneptCharge}), implying a leptonic charge flux asymmetry
\begin{align}
\hspace{-0.2cm}\langle\mathcal{Q}_{\gamma_l}(\vec n)\rangle_{J_Z}=\frac{3}{\pi}\frac{d_{ZZ\gamma_\ell}}{c_{Z,\ell}} \frac{i \epsilon^{i j k} L_{ij} n_k}{L_{ij} \delta^{ij}}=-\frac{A_{FB}^l}{\pi}\cos\theta\,.
\end{align}
Compared with the pure polarization state in Eq.~\eqref{eq:oneptCharge}, the polarization density matrix of the $Z$ current is now determined by the leptonic tensor $L_{ij}$ describing the incoming lepton beams.
Using the SM relation $d_{T_3 T_3 L} = -d_{YYL}$ to simplify our expressions, we have
\begin{align}
A_{FB}^{l}=\frac{3}{2} A_e \frac{\left(1-4 s_W^2\right) d_{T_3T_3 L}}{\left(1-4 s_W^2+8 s_W^4\right)}\,,
\end{align}
where $A_e \cos\theta=i \epsilon^{i j k} L_{ij} n_k/L_{ii}$ is given by the contraction with the leptonic tensor. 
This has a simple interpretation: at tree level, a Z boson decays into two back-to-back leptons. The anomaly manifests as the preferential alignment of the electron with the Z boson spin (experimentally this is converted into an angle with respect to the incoming electron beam). At tree level, the relation between the anomaly and the perturbative calculation of the charge detector from squared tree-level diagrams, shown in \Fig{fig:anomaly_detector}, is also completely transparent.   

The coefficient $A^l_{\text{FB}}$ is commonly referred to as the  ``forward-backward asymmetry"
\begin{align}\label{eq:fb_def}
A^l_{\text{FB}}=\frac{\sigma_F-\sigma_B}{\sigma_F+\sigma_B}\,,
\end{align}
namely the difference in cross section for the electron to be produced in the forward vs. backward direction (defined using the incoming negatively charged lepton).
The leptonic forward-backward asymmetry is one of the most important observables in the history of electron-positron colliders. For early theory work, see \cite{Cabibbo:1961sz,Cung:1972ug,Godine:1972exm,Berends:1973fd,Budny:1974wn}, for a detailed overview of experimental results, see \cite{Wu:1984ik,ALEPH:2005ab,ALEPH:2010aa}, and for a more recent analysis using archival SLD data, see \cite{Cheng:2026zvl}.

For leptonic final states, the ability to directly identify the outgoing electron and positron makes the language of forward-backward asymmetries natural.
The specific relation of leptonic forward-backward asymmetries at the Z-pole to anomalies that we are emphasizing here is therefore at most an amusing re-interpretation. 
However, this simple example illustrates that charge correlator observables capture the rich physics of forward-backward asymmetries, providing the opportunity to use advances from the modern energy correlator program \cite{Moult:2025nhu} to study the physics of forward-backward asymmetries in more complicated final states.
%

\section{Anomalies in Hadronic Fluxes}

Forward-backward asymmetries are also extensively measured on hadronic final states at the Z-pole, where they provide measurements of electroweak couplings to quarks. 
These measurements can be performed inclusively on all hadrons \cite{OPAL:1997tsq,OPAL:1992jsm,ALEPH:1991fba,ALEPH:1996qlh,ALEPH:1998pmr,L3:1991gfs,L3:1998jgx,DELPHI:1991mqi}, on $b$- and $c$-enriched decays \cite{OPAL:1993wua,ALEPH:2001mdb,L3:1992fsb,DELPHI:2004wvq}, and even on $s$-enriched decays \cite{DELPHI:1994aml,SLD:2000jop,DELPHI:1999mkl}. They have also been the focus of intense theoretical efforts starting with the early works of \cite{Djouadi:1994wt,Arbuzov:1991pr,Korner:1985dt,Lampe:1996rt,Jersak:1981sp,Jersak:1979uv,Altarelli:1992fs,Ravindran:1998jw,Catani:1999nf}, and culminating in the complete next-to-next-to-leading order calculation incorporating mass effects \cite{Bernreuther:2016ccf,Bernreuther:2023jgp,Wang:2020ell}. For a detailed review of experimental results, see \cite{ALEPH:2005ab,ALEPH:2010aa}.

In high-multiplicity final states, the concept of a forward-backward asymmetry is much less natural, and relies on the ability to associate multi-particle states with an underlying primary quark direction and charge. This is often done using jets, or thrust hemispheres, combined with measurements of jet charge. While jet charge retains some correlation with the primary quark charge, this association will become worse at future higher-energy colliders which achieve higher multiplicities. Furthermore, these measurements are often not infrared and collinear safe \cite{Catani:1999nf,Weinzierl:2006yt}, complicating the comparison with theoretical calculations. For a recent discussion of QCD uncertainties, see  \cite{dEnterria:2018jsx}. Hadronic asymmetry measurements give rise to one of the longest-standing tensions in precision electroweak fits \cite{Baak:2014ora,Haller:2018nnx,Fischer:2026bka}, which is important to resolve to constrain possible new physics scenarios.

In the study of energy flux in high-multiplicity final states, energy correlators provide the natural field theoretic generalization of the S-matrix, and have proven powerful in connecting theory and experiment \cite{Moult:2025nhu}.
Here we argue that the natural field theoretic generalization of the forward-backward asymmetry to high multiplicity final states is the one-point charge correlator of the hadronic analog of the electromagnetic charge detector, $\mathcal{Q}_{\gamma_B}(\vec n)$ \footnote{We highlight a related early proposal \cite{Nason:1994ad}, which unfortunately does not seem to have gained traction.}. Indeed, this is most beautifully illustrated in the case of conformal field theories discussed above, where despite the fact that there are no quasi-particles to measure as going forward or backward, the asymmetry can be directly measured from the angular distribution of charge flux, and cleanly related to the underlying anomaly. Additionally, we will see that the direct relation between the charge correlator and the anomaly largely protects it from perturbative corrections. It is in the context of hadronic collisions that recognizing anomalies as the underlying physics of forward–backward asymmetries proves most powerful. Charge correlators have also recently been used to characterize other aspects of hadronic fluxes in~\cite{Cao:2026fzq,Monni:2025zyv,Riembau:2024tom}.

In contrast to forward-backward asymmetries, charge flux correlators generalize naturally to many-particle states, and avoid the necessity of jet or hemisphere definitions, and their ambiguous association with primary quarks. 
Using \Eq{eq:chargedet} we can define detector operators for different hadronic currents. Fluxes which are experimentally realizable include $J_f =\bar f\gamma^\mu f$ where $f=b,c$ denotes a heavy flavor quark (recent advances in strange tagging may enable extensions to strange quarks \cite{Defranchis:2026wyw}), and electromagnetically charged hadrons, $J_{\gamma_B}$. The measurement of the complete baryon number current  is challenging experimentally due to the requirement to detect both neutrons and anti-neutrons. For the experimental data analysis in this \emph{Letter}, we will focus on the simplest case of inclusive electromagnetically charged hadrons, $J_{\gamma_B}$. The one-point function of these detectors is infrared and collinear safe. This provides a baseline for the generalization to flavor-exclusive currents.

As for the leptonic current, $J_{\gamma_l}$, we can express the mixed anomaly coefficient for the electromagnetic charge on baryons, $J_{\gamma_B}$, as a linear combination of the anomaly coefficients for $T_3$, $Y$, and $B$
\begin{align}
\label{eq:dZZgammaB}
d_{ZZ\gamma_B} &=\frac{1-4 s_W^2}{2} d_{T_3 T_3 B}+s_W^4\left(d_{T_3 T_3 B}+d_{Y Y B}\right)\,.
\end{align}
Its measurement therefore provides direct access to mixed electroweak anomalies involving baryon number. The non-anomalous nature of the electromagnetic current implies
\begin{align}
d_{ZZ\gamma_B}+ d_{ZZ\gamma_L} =d_{ZZ\gamma}=0\,,
\end{align}
which provides a relation between leptonic and hadronic charge fluxes. 

Unlike in the leptonic sector where we could neglect interactions, QCD corrections violate conformal symmetry and must be carefully incorporated. Remarkably, due to special properties of the anomaly, corrections are highly constrained, preserving a relation between the anomaly coefficient and the one-point charge correlator. Due to the asymptotic freedom of QCD, at Z-pole energies, the breaking of conformal symmetry in QCD can be treated perturbatively.  For simplicity, we ignore explicit breaking arising from Yukawa couplings (masses). We have verified that such corrections are small. Perturbative QCD corrections are flavor universal, and therefore apply to any hadronic current, so we keep our discussion generic. Ignoring explicit breaking, the three-point function of currents, $\left\langle J_{ \alpha}(x) J_{\beta}(y) J_{\gamma}(z)\right\rangle$, satisfies the anomalous conformal Ward identities, whose solution is \cite{Braun:2003rp} 
\begin{align}
\left\langle J_{ \alpha}(x) J_{\beta}(y) J_{\gamma}(z)\right\rangle&=\Delta_{\alpha \beta \gamma}(x-y,z-y)   \\
& +\beta(g_i) \bar \Delta_{\alpha \beta \gamma} (x-y,z-y;g_i)\,. \nn
\end{align}
Here $\Delta_{\alpha \beta \gamma}(x-y,z-y) $ is fixed by the anomaly, while  $\bar \Delta_{\alpha \beta \gamma} (x-y,z-y;g_i)$ captures non-conformal contributions, and appears with an explicit prefactor of the $\beta$-function.
 While this result applies generically, here we focus on the dominant QCD corrections, so that the $\beta$-function is that of QCD. Explicit calculations of the three-point correlator in QCD \cite{Melnikov:2006qb,Mondejar:2012sz} have verified this structure. We see that the relation between the anomaly coefficient and the charge correlator distribution is only modified by $\beta$-function terms, which start at $\alpha_s^2$. Combining this with the perturbative results of \cite{Rijken:1996npa}, we can explicitly compute the result for the correlator of electromagnetic flux on baryons in QCD
\begin{align}\label{eq:onept_QCD}
\langle  \mathcal{Q}_{\gamma_B}(\vec{n})\rangle_{J_Z} &=  A_e\frac{3}{\pi}\frac{d_{ZZ\gamma_B}}{c_{Z,B}} \cos\theta \\
&\times \left(1-\left(\frac{\alpha_s}{4\pi}\right)^2 12 \beta_0C_F\zeta_3 +\mathcal{O}\left(\alpha_s^3\right)\right) \,,\nn
\end{align}
where $\beta_0=11/3C_A-2/3 n_f$ is the one-loop $\beta$-function.
Due to the protected structure of the anomaly, perturbative corrections first appear at $\alpha_s^2$ and are therefore extremely tiny. We are not aware of other QCD event shape observables where the one-loop correction vanishes. In our numerical results, we will also incorporate the $\alpha_s^3$ corrections, obtained from  \cite{Rijken:1996npa}, although they are lengthy, so we do not reproduce them here.
 Additionally, there are non-perturbative corrections. The charge sum rule used to connect the flux measured on hadrons to the underlying partonic calculation receives corrections from soft hadrons that are not suppressed and are found to be of order $10\%$ in parton shower simulations, from which we also extract their value. We plan to characterize these corrections in more detail in future work.

This analysis establishes the theoretical interpretation of charge flux measurements in the SM, and their relation to its underlying anomaly structure.  Remarkably, despite the complexity of a hadronic QCD collision, the anomaly is cleanly imprinted in the macroscopic structure of charge flux.
The use of the charge correlator avoids the necessity of jets, hemispheres, and primary quarks, and directly extracts the asymmetry, \Eq{eq:onept_QCD}, from a simple measurement.
To verify that this is borne out in data, we use archival DELPHI data to measure the one-point charge correlator of charge flux on hadrons, and compare with our theoretical predictions in \Eq{eq:onept_QCD}.

\section{Archival Analysis with DELPHI Data}

\begin{figure}[t]
\centering
\begin{tikzpicture}
  \node[anchor=south west,inner sep=0pt] (base) at (0,0)
    {\includegraphics[width=\figthreewidth]{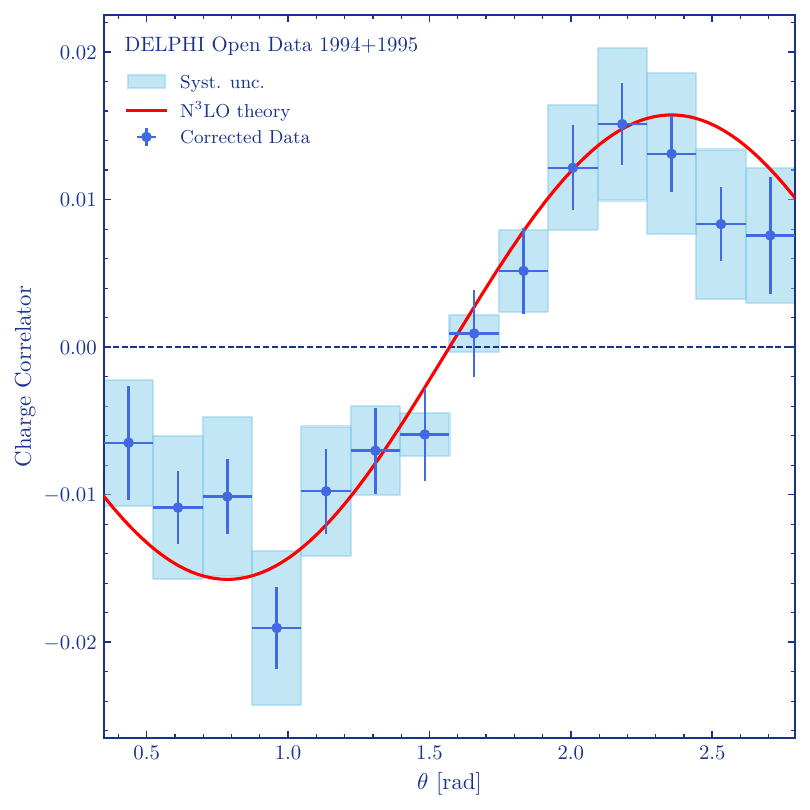}};
  \begin{scope}[x={(base.south east)},y={(base.north west)}]
    \node[anchor=north west,inner sep=0pt] at (0.571,0.485)
      {\includegraphics[width=0.400\figthreewidth]{figs/charge_asym_color.pdf}};
    \node[anchor=north west,inner sep=0pt] at (0.395,0.922)
      {\includegraphics[width=0.117\figthreewidth,
                        trim=129bp 53bp 206bp 45bp,clip]{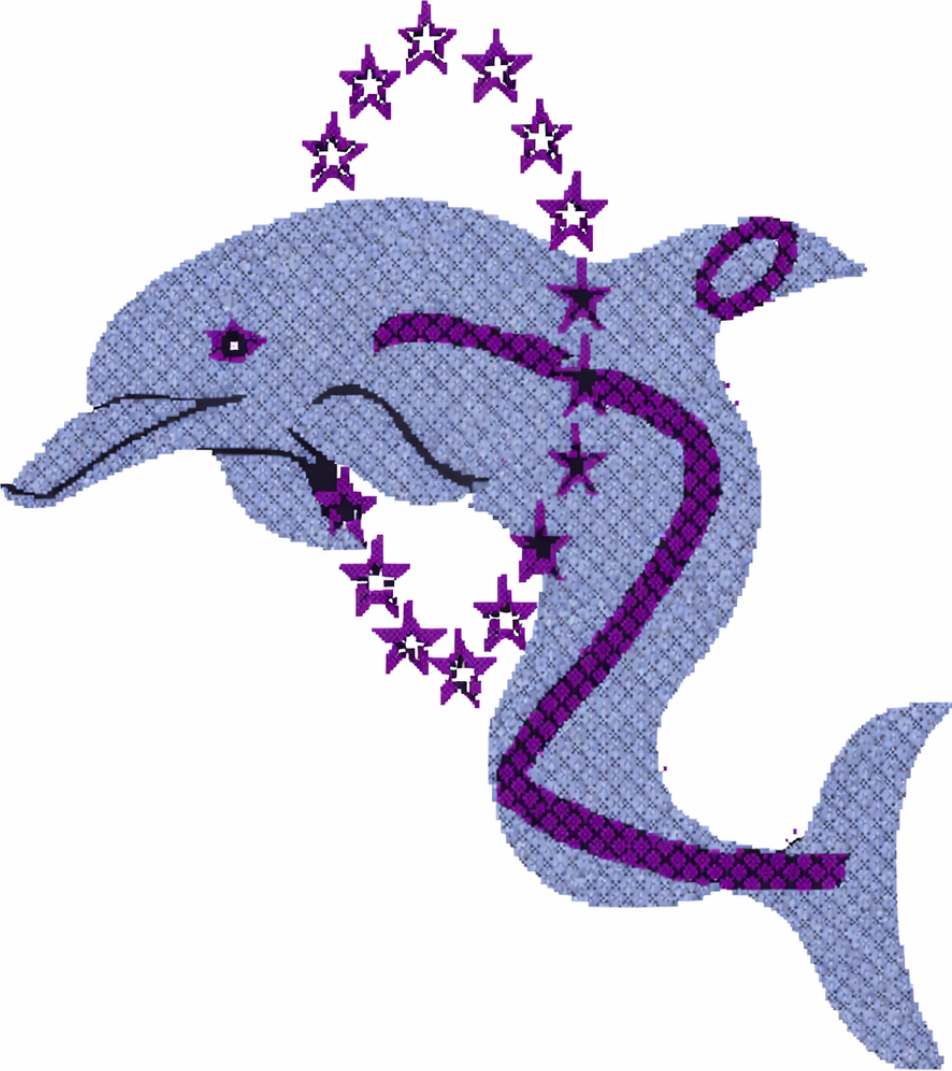}};
    \node[anchor=north west,inner sep=0pt] at (0.530,0.913)
      {\includegraphics[width=0.116\figthreewidth]{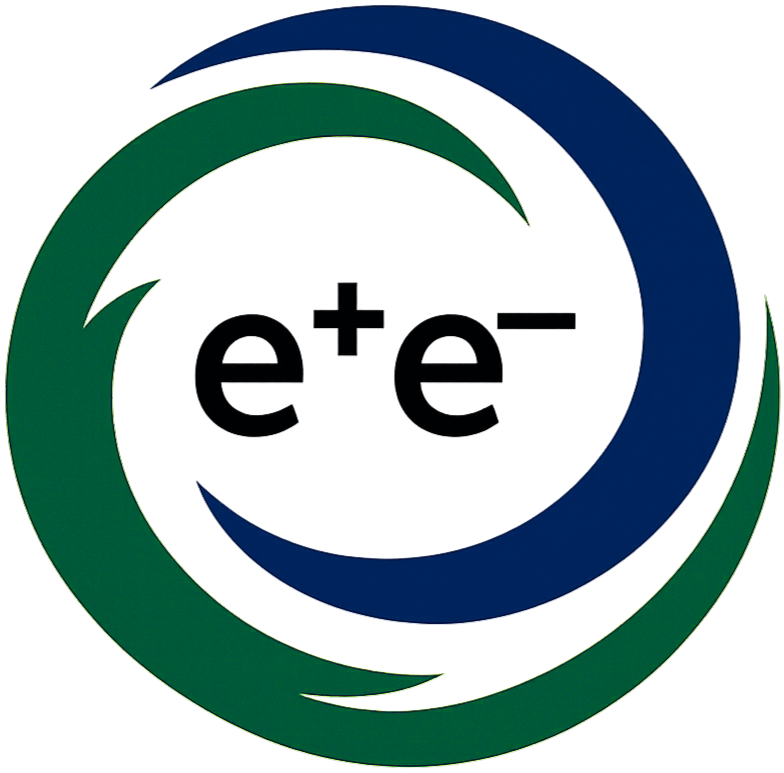}};
  \end{scope}
\end{tikzpicture}
\caption{The electromagnetic charge flow on hadrons $\langle Q_{\gamma B} (\vec{n}) \rangle_{J_Z}$ measured in $e^+e^- \to {\rm hadrons}$ at $\sqrt{s}=M_Z$ using archival DELPHI data. The $\theta$ odd modulation is caused by the underlying mixed electroweak anomaly of this current.}
\label{fig:data}
\end{figure}

In recent years, there has been an effort to reanalyze~\cite{Electron-PositronAlliance:2019cpi, Electron-PositronAlliance:2021kig,Electron-PositronAlliance:2023klx,Electron-PositronAlliance:2025hze, Electron-PositronAlliance:2025fhk} archival LEP data using modern experimental techniques. To probe electromagnetic charge flow in hadrons, we use data recorded by the DELPHI detector~\cite{DELPHI:1990cdc, DELPHI:1995dsm} at the $Z$ pole, $\sqrt{s}=91.2$~GeV, during the 1994 and 1995 LEP runs and released through the DELPHI Open Data program~\cite{DELPHI:2024opendata, DELPHI:2024policy}. These data comprise an integrated luminosity of $46$~pb$^{-1}$ from 1994 and $15$~pb$^{-1}$ of on-peak data from the 1995 energy scan, for a total of $61$~pb$^{-1}$, from which approximately $1.6$~million hadronic $Z$ decays are selected. To model detector effects, $5$~million $Z^0\to q\bar{q}$ events per year are generated with \textsc{Pythia}~8.3~\cite{Bierlich:2022pfr} using the Monash 2013 tune~\cite{Skands:2014pea} and processed through the restored DELSIM detector simulation chain~\cite{DELSIM}. \textsc{KK}2f~\cite{Jadach:1999vf}$+$\textsc{Pythia}~6.1~\cite{Sjostrand:2000wi} samples from the official DELPHI production~\cite{DELPHI:OpenData:kk2f_pythia_94, DELPHI:OpenData:kk2f_pythia_95}, totaling fewer than $2.5$~million events across the two years, are used to assess generator-model dependence.

The one-point charge correlator is constructed in data as $\mathcal{Q}_{\gamma_B}(\theta)=n^+(\theta)-n^-(\theta)$, the difference between the positive- and negative-track density as a function of the polar angle $\theta$ with respect to the $e^{-}$ beam direction. Charged tracks are selected, following established DELPHI selections~\cite{DELPHI:2003yqh, DELPHI:2004wvq, Zhang:2025delphiEEC}. Charged tracks are required to fall within the polar-angle acceptance $20^\circ\le\theta\le160^\circ$, to have a measured length above $30$~cm and a relative momentum uncertainty $\Delta p/p\le1.0$. Two requirements are tightened to reduce charge misreconstruction: the impact parameters are restricted to $|d_0|\le0.6$~cm and $|z_0|\le1.0$~cm to reject mismeasured tracks at large impact parameters, and the transverse momentum is raised to $p_{\rm T}>2$~GeV to suppress soft tracks that are sensitive to charge mismodeling from secondary nuclear interactions in the detector material. Hadronic events are selected by requiring at least seven good charged tracks, total reconstructed energy $E_{\rm tot}\ge0.5\,E_{\rm cm}$, and a thrust-axis polar angle $30^\circ\le\theta_{\rm thrust}\le150^\circ$. The requirement on the total reconstructed energy additionally suppresses radiative-return events with hard initial-state radiation, for which the effective collision energy lies well below the $Z$ pole and QED processes reintroduce $\theta$-odd contributions unrelated to the on-pole anomaly signal.

The characteristic $\theta$-odd modulation is already clearly visible in the reconstructed data, before any charge-dependent correction. A charge-dependent tracking bias with a parity-odd angular dependence would, however, produce a modulation of the same form. A dedicated correction is therefore applied to establish that the observed signal is physical rather than instrumental. Simulation studies indicate that the charge misreconstruction is driven by single-track-level detector effects, i.e., geometry, material description, and alignment, rather than by the event environment or event flavor composition. The measured charge correlator is therefore corrected for detector effects in two stages. First, a simulation-based correction is derived and applied separately for positive and negative charged tracks. Second, the residual data--simulation charge reconstruction differences are bounded by a tag-and-probe approach in the single-prong $e^+e^-\to Z\to\tau^+\tau^-$ sample, where charge conservation fixes the true charge of the probe track. The resulting residual, measured in bins of probe $p_{\rm T}$ and $\theta$, is then transferred to the hadronic events. The parity-odd component of the residual, which is aligned with the angular structure of the signal, is statistically consistent with zero but is propagated in full. The data-driven charge calibration procedure therefore gives the dominant systematic uncertainty, whose size is set by the statistics of the control sample. A detailed description of all analysis steps is provided in a companion experimental paper and analysis note.

\section{The One-Point Charge Correlator in Hadronic Z-decays}

In \Fig{fig:data} we present our measurement of the one-point correlator of electromagnetic flux on hadrons $\langle \mathcal{Q}_{\gamma_B} (\vec{n}) \rangle_{J_Z}$ in $e^+e^- \to {\rm hadrons}$ at $\sqrt{s}=M_Z$ using archival DELPHI data. The characteristic $\theta$-odd modulation is cleanly visible in the data, consistent with the N$^3$LO QCD prediction of \Eq{eq:onept_QCD}, which takes the form $\sin\theta\cos\theta$ once the Jacobian of the experimental $\theta$ binning is included. This is a direct laboratory observation of a macroscopic consequence of the electroweak anomalies of \Eq{eq:Banom}. In comparing the measurement to theory, we have verified using parton shower simulations that the experimental cut, $p_{\rm T}>2$~GeV, and residual soft initial-state radiation near the $Z$ pole each have a small effect, comparable with current experimental uncertainties, on the distribution. Non-perturbative corrections are extracted from parton shower simulations and incorporated into the theory prediction. We have also verified that $b$-quark mass effects are negligible. The uncertainties on the theoretical calculation in \Fig{fig:data} arise only from perturbative scale variation, and are tiny since the perturbative corrections start at $\mathcal{O}(\alpha_s^2)$.

While energy correlators have now been extensively measured, see in particular the ALEPH re-analysis \cite{Electron-PositronAlliance:2025fhk}, this \emph{Letter} illustrates the experimental feasibility of measuring charge correlators. As emphasized above, formulating the physics of forward-backward asymmetries in terms of charge correlators offers many theoretical advantages. With a first measurement of the charge correlator and an understanding of their subtleties, we take the first step towards making their application to precision electroweak physics a reality. 

Additionally, we believe that the charge correlator measurement in \Fig{fig:data} provides a beautiful illustration of a macroscopic consequence of the electroweak anomaly equation \Eq{eq:Banom} in the laboratory. While inclusive hadronic charge asymmetries have been measured at the Z-pole for decades \cite{OPAL:1997tsq,OPAL:1992jsm,ALEPH:1991fba,ALEPH:1996qlh,ALEPH:1998pmr,L3:1991gfs,L3:1998jgx,DELPHI:1991mqi}, the direct connection to the anomaly provided by the charge correlator is pleasing, and we believe will lead to improvements in its theoretical description.

\section{Conclusions}
In this \emph{Letter}, we have emphasized that charge asymmetries at the $Z$-pole are macroscopic manifestations of electroweak anomalies. 
We proposed a new way to measure these asymmetries using charge correlators, which directly connects the experimental observable with the underlying field theory description. 
Building on the case of conformal field theories, in which the magnitude of the angular asymmetry is exactly fixed by the anomaly coefficient, we used anomalous conformal Ward identities to show that this relation is preserved in the SM up to perturbative corrections that can be reliably computed at $Z$-pole energies.
This provides a new formulation of forward-backward asymmetry measurements, identical for leptonic final states, but presenting several theoretical and experimental advantages for many-body hadronic final states.
Additionally, it initiates an exciting connection between recent progress in the energy correlator program \cite{Moult:2025nhu}, anomalies,  and electroweak physics.
Using archival DELPHI data, we performed a first measurement of the one-point correlator of electromagnetic flux on hadrons $\langle Q_{\gamma_B} (\vec{n}) \rangle_{J_Z}$ in  $e^+e^- \to {\rm hadrons}$ at $\sqrt{s}=M_Z$. 
The measured asymmetry agrees well with our theoretical expectation, illustrating the feasibility of this approach for studies of hadronic asymmetries at the Z-pole.

The approach proposed in this \emph{Letter} generalizes straightforwardly to asymmetries of other hadronic currents, such as heavy flavor currents, as well as to the inclusion of polarization or chirality information \cite{Barata:2026eth}. Particularly for heavy flavors, the application of modern flavor taggers to archival data \cite{Defranchis:2026wyw} should provide significant improvements compared to studies at LEP/SLC.
It will be exciting to pursue such measurements using archival LEP and SLC data.
Looking further ahead, the proposed FCC-ee~\cite{FCC:2018evy} and CEPC~\cite{CEPCStudyGroup:2018ghi} colliders will produce orders of magnitude more $Z$ decays than LEP, with corresponding improvements in statistical precision. 
Such measurements may help resolve the longstanding tension in precision electroweak observables and provide new insights into the electroweak sector of the SM.

\emph{Acknowledgements.}---We thank Yang Bai, Nathaniel Craig, Shirley Li, Carlos Wagner for useful discussions. We thank Aneesh Manohar, Michael Peskin, Riccardo Rattazzi, Marc Riembau, and Paolo Nason for useful comments on the manuscript. 
We are particularly grateful to David B. Kaplan for helpful discussions on the interpretation of our results in terms of anomalies in the SM, which contributed significantly to the presentation. 
We thank Michael Peskin for help with the early literature on forward-backward asymmetries.
We thank Xiaoyuan Zhang for help with the early Russian literature on electron-positron colliders.
We are grateful to the DELPHI Collaboration and the DELPHI data-preservation team for making these data publicly accessible, and we thank Dietrich Liko and Ulrich Schwickerath for their guidance on the DELPHI Open Data.
We thank the KITP Santa Barbara for hospitality while this work was initiated.
K.L. is supported by the U.S. Department of Energy under contract DE-AC02-06CH11357.
I.M. is supported by the DOE Early Career Award DE-SC0025581 and the Sloan Foundation.
J.Z. is supported in part by Vanderbilt University faculty funds of Y.C. Y.-J. Lee is supported by the Department of Energy, Office of Science, under Grant No. DE-SC0011088.

\bibliography{refs}

\end{document}